\documentclass[prl,twocolumn,showpacs]{revtex4}

\usepackage{amssymb,amsmath,bm,epsfig,graphics,subfigure}

\begin{document}

\title{Landau Quantization in Twisted Bilayer Graphenes: the Dirac Comb}
\author{M. Kindermann$^1$ and E.J. Mele$^2$}
    \email{mele@physics.upenn.edu}
    \affiliation{$^1$School of Physics Georgia Institute of Technology Atlanta, GA 30332 \\
    $^2$Department of Physics and Astronomy University of Pennsylvania Philadelphia PA 19104  \\}
\date{\today}

\begin{abstract}
We study the Landau quantization of the electronic spectrum for graphene bilayers that are rotationally faulted to produce
periodic superlattices. Commensurate twisted bilayers exist in two families distinguished by their sublattice exchange parity.
We show that these two families exhibit distinct Landau quantized spectra distinguished both by the interlayer coupling of
their zero modes and by an amplitude modulation of their spectra at energies above their low energy interlayer coherence
scales. These modulations can provide a powerful experimental probe of the magnitude of a weak coherence splitting in a bilayer
and its low energy mass structure.
\end{abstract}

\pacs{73.22.Pr, 77.55.Px, 73.20.-r} \maketitle

The massless Dirac model that describes the low energy electronic physics of single layer graphene is generally preempted by
the interlayer motion of the charge carriers in few layer graphenes (FLG's) \cite{MM }. This physics is well appreciated for
Bernal ($AB$) stacked bilayer graphenes where the  interlayer coherence scale is $\sim 0.2$ eV and the low energy electronic
physics \cite{McCann ,Ohta ,Castro ,Li } is readily distinguishable from that of single layer graphene \cite{Novoselov ,Zhang
}. Surprisingly, one finds only relatively weak (if any) effects of coherent interlayer motion in rotationally faulted
(twisted) FLG's where the crystallographic axes in neighboring layers are rotated by angles $\theta \neq n \pi/3$ \cite{Berger
,Hass2,LiAndrei ,Sprinkle ,Hicks ,GeimAndrei}. Theory suggests that the interlayer coherence scale is suppressed in these
structures \cite{LpD ,Hass2,Latil ,Shallcross ,GTdL } but {\it below} this scale one confronts a commensuration problem where
the electronic physics is controlled both by the rotation angle and the atomic registry of neighboring twisted layers
\cite{MeleRC,Bistritzer}. The low energy electronic properties of these structures are very rich but they are not yet well
understood \cite{Hicks ,BMNandV}.

In this Letter we show that the interlayer coupling in twisted bilayers produces novel fingerprints in their Landau level
spectra. We show that small differences in the Landau level spectra from two overlapping weakly coherence-split bands generate
an interference pattern detectable as an amplitude modulation of the spectrum: the ``Dirac comb." The period of the modulation
can greatly exceed the underlying (presumably small) coherence scale and its phase contains information about the low energy
mass structure. This opens the possibility of probing even weak interlayer coherence in these systems using spectroscopy above
their interlayer coherence scales.

Landau quantization of a twisted bilayer is studied using a long wavelength theory that includes the effects of the
 lattice
commensurability \cite{MeleRC}. We study commensurate rotational faults where we fix overlapping $A$-sublattice sites at the
origin of the $AA$ stacked stucture and rotate one layer through angle $\theta$ with respect to the other, generating a new
commensurate superlattice. Commensurate rotations can be classified according to their sublattice parity: sublattice even (SE)
structures are invariant under exchange of sublattice labels $A$ and $B$ while sublattice odd (SO) structures break this
sublattice symmetry. In this notation $AA$ stacking (all sites in neighboring layers eclipsed) has $\theta =0$ (SE), Bernal
$AB$ stacking has $\theta = \pi/3$ (SO). Sublattice parity can be used to classify {\it any} commensurate rotated structure and
determines the form of the allowed momentum-conserving interlayer terms in the low energy Hamiltonian \cite{MeleRC}. Each layer
has Dirac cones in two valleys centered on the $K$ and $K'$ points described by the (unrotated) long wavelength Hamiltonians
\cite{DPD }
\begin{eqnarray}
H_K = - i \hbar v_F \sigma \cdot \nabla; \,\, H_{K'} = \sigma_y H_K \sigma_y,
\end{eqnarray}
where the  $2 \times 2$ $\sigma$ matrices act on the sublattice (pseudospin) degrees of freedom. For the SE structures the
interlayer motion is valley preserving and is described by the interlayer operator
\begin{eqnarray}
H_{\rm int}^{\rm SE} = {\cal V} e^{i \vartheta} \exp \left(i \phi \sigma_z \tau_z \right),
\end{eqnarray}
where the $2 \times 2$ $\tau$ matrices act on the valley indices. For the SO structures the coupling is both interlayer {\it
and} intervalley and is represented by the interlayer matrix
\begin{eqnarray}
H_{\rm int}^{\rm SO} = \frac{{\cal V}}{2} e^{i \vartheta} \left(1 +\sigma_z \right) \tau_x
\end{eqnarray}
(or its valley reversed partner $\sigma_x H_{\rm int}^{\rm SO} \sigma_x$). In Eqns. (2) and (3) the coupling strength ${\cal
V}$, and the pseudospin rotation angle $\phi$ are determined by the details of the microscopic Hamiltonian while the overall
phase $\vartheta$ can be removed by a $U(1)$  gauge transformation.

\begin{figure}
\begin{center}
  \includegraphics[angle=0,width=\columnwidth,bb=61 176 590 729]{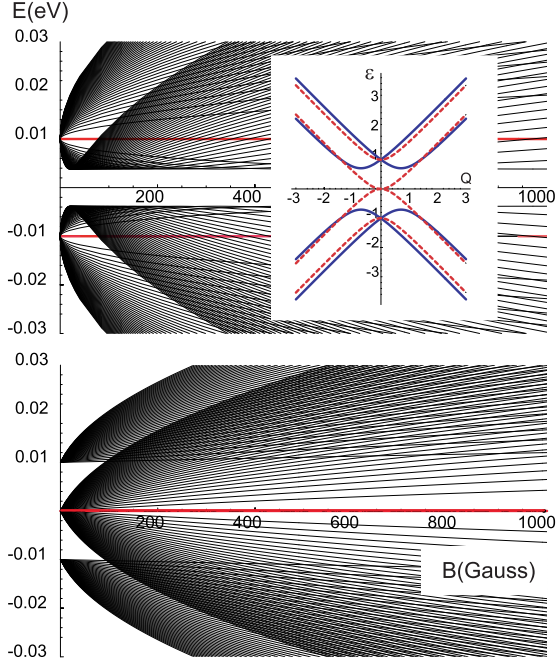}
  \caption{\label{varyfield}  Inset: Dispersion of the coherence-split low energy bands  for commensurate twisted graphene bilayers that are even (blue, solid) and odd
  (red, dashed) under sublattice exchange. The plot gives the dimensionless energy $\varepsilon = E/{\cal V}$ as
  a function of the scaled momentum $Q=\hbar v_F q/{\cal V}$ where ${\cal V}$ is the interlayer coherence scale and the pseudospin rotation parameter $\phi=\pi/4$
  for the SE structure. Main figure: Landau level spectra Eqns. (\ref{SEspec}) and (\ref{SOspec}) for twisted bilayers for SE and SO families as a function of the
  field strength with coherence scale ${\cal V} = 10 \, {\rm meV}$. The red lines are states spanned by the zero modes of the two layers. }
\end{center}
\end{figure}

The low energy dispersion relations for these two families are shown in  the inset of Figure \ref{varyfield}. The spectra of SO
bilayers have a pair of bands gapped by the interlayer coupling and a contact point between two quadratically dispersing low
energy bands, analogous to the situation for the Bernal stacked bilayer except for a $\theta$-dependent interlayer coherence
scale ${\cal V}$ which can be significantly reduced. The SE structures feature a {\it pair} of Dirac nodes at $q=0$ with equal
weights on the two layers and a coherence splitting produced by their interlayer coupling. Note that the SE bilayers are
generically gapped systems (their band extrema occur on a ring in reciprocal space) due to an avoided crossing of its two
coherence-split branches. The size of this gap is determined by the pseudospin rotation angle $\phi$ in Eqn. (2) and it vanishes
for the special case of $AA$ stacking where $\phi=0$ by symmetry.

To study this system in a perpendicular magnetic field $\vec B$ we introduce a vector potential in the symmetric gauge $\vec A
= (\vec B \times \vec r)/2$. Defining $z = (x + iy)/\ell_B$ with magnetic length $\ell_B = \sqrt{\hbar/e B}$ and cyclotron
frequency $\omega_c = \sqrt{2} v_F/\ell_B$ the Hamiltonian at the $K$ point of the $j$-th layer ($j=1,2$) is transformed to a
Landau level (LL) basis
\begin{eqnarray}
h_{K,j} = - i \omega_c \left(%
\begin{array}{cc}
  0 & a e^{-i \theta_j} \\
  -a^\dagger e^{i \theta_j} & 0 \\
\end{array}%
\right)
\end{eqnarray}
where $a^\dagger=\left(-2 \partial_{\bar z} +z/2 \right)/\sqrt{2}$ is the Landau level raising operator and $\theta_j$ is the
rotation angle of the $j$-th layer. The corresponding Hamiltonian at the $K'$ point is $h_{K',j} = \sigma_y h_{K,j} \sigma_y$.
Since the interlayer coupling matrices in Eqns. (2) and (3) are {\it local} in the layer-projected coordinates the bilayer LL
Hamiltonian has the form
\begin{eqnarray}
{\cal H}_{KK}^{\rm SE} = \left(%
\begin{array}{cc}
  h_{K,1} & {\cal V}  \exp \left(i \phi \sigma_z  \right) \\
  {\cal V}  \exp \left(-i \phi \sigma_z  \right) & h_{K,2} \\
\end{array}%
\right) \label{SEbilayerham}
\end{eqnarray}
for SE bilayers and in the SO case
\begin{eqnarray}
{\cal H}_{KK'}^{\rm SO} = \left(%
\begin{array}{cc}
  h_{K,1} & {\cal V}  \left(1 +\sigma_z \right)/2 \\
 {\cal V}  \left(1 +\sigma_z \right)/2 & h_{K',2} \\
\end{array}%
\right). \label{SObilayerham}
\end{eqnarray}
 Particle-hole symmetry allows the spectrum to be studied by squaring ${\cal H}$, yielding for SE bilayers
 \begin{widetext}
\begin{eqnarray}
\left ( {\cal H}_{KK}^{\rm SE} \right)^2 =
 \left(%
\begin{array}{cccc}
  \omega_c^2 (a^\dagger a + 1) + {\cal V}^2 & 0 & 0 & -i \omega_c {\cal V} (ae^{-i\alpha_1} +a e^{-i \alpha_2}) \\
 0 &\omega_c^2 a^\dagger a  + {\cal V}^2&i \omega_c {\cal V} (a^\dagger e^{i\alpha_1} +a^\dagger e^{i\alpha_2})&0 \\
 0&-i \omega_c {\cal V} (a e^{-i\alpha_1} +a e^{-i\alpha_2})&\omega_c^2 (a^\dagger a +1)  + {\cal V}^2&0 \\
i \omega_c {\cal V} (a^\dagger e^{i\alpha_1} +a^\dagger e^{i\alpha_2})&0&0& \omega_c^2 a^\dagger a + {\cal V}^2
\end{array}%
\right) \label{SE4by4}
\end{eqnarray}
\end{widetext}
where $\alpha_1 = \theta_1 +\phi$ and $\alpha_2 = \theta_2 - \phi$ and for the SO bilayer
\begin{widetext}
\begin{eqnarray}
\left ( {\cal H}_{KK'}^{\rm SO} \right)^2 =
 \left(%
\begin{array}{cccc}
  \omega_c^2 (a^\dagger a + 1) + {\cal V}^2 & 0 & 0 & -i \omega_c {\cal V} a^\dagger e^{i \theta_2}   \\
 0 &\omega_c^2 a^\dagger a   &i \omega_c {\cal V} a^\dagger e^{i \theta_1} &0 \\
 0&-i \omega_c {\cal V} a e^{-i \theta_1} &\omega_c^2 a^\dagger a +{\cal V}^2  &0 \\
i \omega_c {\cal V} a e^{-i \theta_2} &0&0& \omega_c^2 (a^\dagger a +1)
\end{array}%
\right) \label{SO4by4}
\end{eqnarray}
\end{widetext}
along with their valley reversed partners. Eqns. (\ref{SE4by4}) and (\ref{SO4by4}) demonstrate that for either family  of
structures the Hamiltonian is block diagonal in a particular layer- and valley-polarized Landau level basis. For SE structures
the interlayer coupling scatters states from LL$_n$ in one layer only into   states from LL$_{\pm n}$ in the same valley of
the neighboring layer, while in the SO structure LL$_n$ in valley $K$ of layer 1 couple only to LL$_{\pm (n-1)}$ in valley $K'$
of layer 2. Within these subspaces the Hamiltonians (\ref{SE4by4}) and (\ref{SO4by4}) can be diagonalized yielding the
 spectra
\begin{eqnarray}
\varepsilon^{\rm SE}_{n,\kappa,\nu} = \pm \sqrt{n \omega_c^2 + {\cal V}^2 + \kappa {\cal V} \sqrt{2n \omega_c^2(1 + \cos
\phi)}} \label{SEspec}
\end{eqnarray}
and
\begin{widetext}
\begin{eqnarray}
\varepsilon^{\rm SO}_{n,\kappa,\nu} =
 \pm  \sqrt{\frac{(2n+1) \omega_c^2 + {\cal V}^2  +\kappa \sqrt{\omega_c^4 +
2(2n+1)\omega_c^2 {\cal V}^2 + {\cal V}^4}}{2}},  \label{SOspec}
\end{eqnarray}
\end{widetext}
with valley ($\nu = \pm 1$) and branch ($\kappa = \pm 1$) indices,  setting $\hbar = 1$. The eigenstates of the SE structures come in valley-degenerate pairs. SO structures have degenerate pairs of states along with
a quartet of orbital states at $\varepsilon=0$ (with indices $n=0$, $\kappa = -1$, and $\nu =\pm 1$).

In Figure \ref{varyfield} we plot the LL spectra of Eqns. (\ref{SEspec}) and (\ref{SOspec}) as a function of the field
strength. For SE structures, the states disperse $\propto \sqrt{B}$ away from the two coherence-split Dirac nodes and evolve at
high field into two pairs of branches that disperse away from a common charge neutrality level. For SO structures the fermions
are massive in both the low energy (around $E=0$) and high energy (around $E=\pm {\cal V}$) branches and the LL dispersion is
thus linear in $B$ when $\omega_c \ll {\cal V}$ crossing over to the expected the $\sqrt{B}$ dependence in high field. For
Bernal stacking ${\cal V} \gg \omega_c$ at experimentally realizable fields and one can  integrate out the high energy band to
obtain an effective two band model for the ``weak field" limit of the SO spectra \cite{McCann }. By contrast in twisted SE and
SO structures the coherence scale collapses and all the degrees of freedom become accessible.

A striking prediction of Eqns. (\ref{SEspec}) and (\ref{SOspec}) is that the overlapping branches of Landau levels produce an
amplitude modulation of the density of states. When $\varepsilon \gg {\cal V}$ for SE structures and $\varepsilon \gg ({\cal
V}, \omega_c^2/{\cal V})$ for SO structures we can approximate $\varepsilon \approx \pm (\sqrt{n} + \kappa \beta {\cal V})$
where for (SE,SO) bilayers $\beta = (\cos(\phi/2),1/2)$. The density of states is  enhanced whenever the energy of LL$_n$ in
one branch ($\kappa =-1$) overlaps the energy of LL$_{n-m}$ in the other ($\kappa = 1$). This occurs for $m=4 \beta {\cal V}
\sqrt{n}/\omega_c$ when $n \gg m$, and produces a beating pattern with period $\Delta E = \omega_c^2/4 \beta {\cal V}$, most
evident when $\Delta E \gg \omega_c \gg {\cal V}$.  The superposition of the LL spectra thus produces a ``Dirac comb."

Scanning tunneling spectroscopy (STS) \cite{STM} in modest magnetic fields can access these quantum oscillations in the single
layer-projected density of states  $\rho_1$ and thus quantify the interlayer coherence. Theoretically,  $\rho_1$ is conveniently studied  by integrating out
the second layer \cite{KindermannFirst} yielding an effective Hamiltonian for the first (exposed) layer $ H_1^{\rm eff}(E) =
H_1 + H_{\rm int} \left(E - H_2 \right)^{-1} H_{\rm int}^\dagger$ containing a state- and energy-dependent self energy that is
evaluated separately in each invariant subspace of Eqns. (\ref{SEbilayerham}) and (\ref{SObilayerham}). The density of states is
then obtained by a trace over the sublattice degrees of freedom and invariant subspaces: $ \rho_1(E) = \sum_{n,\nu} {\rm Im} \,{\rm tr}
\left[E - i \gamma - H_{1,n,\nu}^{\rm eff} (E - i \gamma) \right]^{-1} $ with level broadening $\gamma$. Importantly, in both SE
and SO structures the zero mode of the surface-projected problem occupies a one dimensional subspace with
\begin{eqnarray}
H_{1,n=0}^{{\rm eff,SE}} \!= |{\cal V}|^2/E; \,\, H_{1,n=0,\nu}^{{\rm eff,SO}} \!=  \delta_{\nu,-1} E |{\cal V}|^2/(E^2\! -\!
\omega_c^2). \label{zeromodes}
\end{eqnarray}
Eqn. (\ref{zeromodes}) demonstrates that the zero modes of each layer are always coherence-split by  $2{\cal V}$ in SE bilayers
but remain {\it exactly decoupled} for all SO bilayers.

These features can be seen clearly in the densities of states in Figures (2) and (3). The coherence splitting of the zero modes
for the SE structure can be contrasted with the degenerate layer-polarized zero energy states for the SO structures. At higher
energy (Fig. (3)) quantum oscillations are clearly seen for both families. Although the $B=0$ dispersions are nearly the {\it
same} for these two structures at  energies $E\gg {\cal V}$ (Fig. 1) the amplitude modulations near $E\simeq \omega_c^2/ \beta {\cal V}$ are
phase shifted, reflecting their distinct low energy mass structure.
\begin{figure}
\begin{center}
  \includegraphics[angle=0,width=2.5in,bb=162 299 450 493]{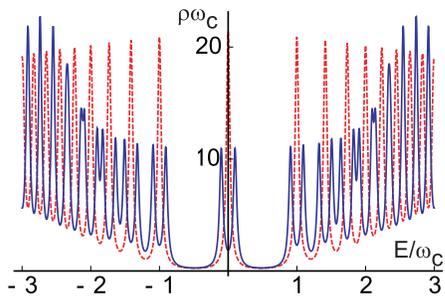}
  \caption{\label{lowenergy} Layer projected density of states for SE (blue) and SO(dashed red) twisted bilayers as a function
  of $E/\omega_c$ ($\hbar =1)$. Data are plotted for ${\cal V}/\omega_c = 0.1$, $\gamma = .03 \times \omega_c$ and $\phi=\pi/4$. The
  low field states of the SO bilayer are weakly coherence split for the SE bilayer.  }
\end{center}
\end{figure}
\begin{figure}
\begin{center}
  \includegraphics[angle=0,width=3.0in,bb=73 156 503 564]{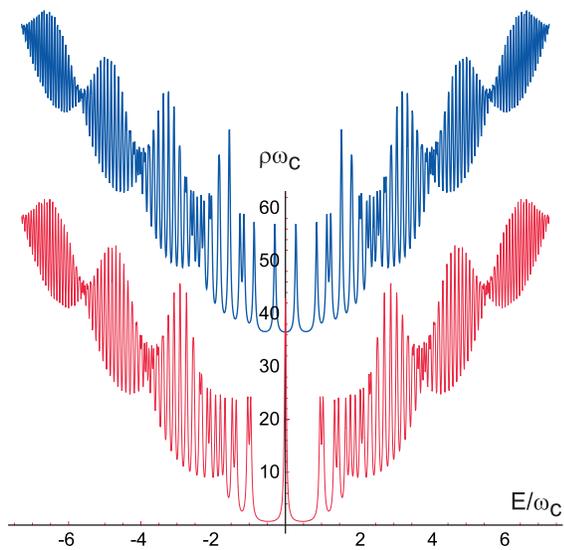}
  \caption{\label{SObeat} Layer projected density of states as a function of $E/\omega_c$ over a wide energy
  range showing the amplitude modulation in the Dirac comb. Data are plotted for ${\cal V}/\omega_c = 0.3$ and $\phi=2\pi/3$. Top panel(blue, vertically offset) is for an SE bilayer
  and the lower panel (red) is for an SO bilayer.}
\end{center}
\end{figure}

The modulations reflect the atomic registry in a commensurate bilayer and are most pronounced for short period superlattices.
They are observable for $B> {\cal V}^2/2 e \hbar v_F^2$, i.e. at a field scale of only $0.1 \, {\rm T}$  for ${\cal V} \sim 10
\, {\rm meV}$. For small fault angles the coherence scale ${\cal V}$ collapses, the magnetic field scale is correspondingly
reduced and measurements will ultimately be limited by the finite quasiparticle lifetime.  Interestingly, the modulation period
$\Delta E = \hbar^2 \omega_c^2/4 \beta {\cal V} = \hbar e B/M^*$ describes the cyclotron frequency of a {\it massive} particle
with $M^* \sim 3.5 \times 10^{-3} m_{\rm e}$ when ${\cal V} = 10 \, {\rm meV}$. This can be understood semiclassically by
noting that the quantization conditions for orbits in the two weakly coherence-split Dirac bands acquire the difference of one
quantum from one maximum of the amplitude modulation to the next. An effective quantization condition within the phase space
annulus bounded by the two coherence-split  bands thus follows.

These low field quantum oscillations are physically distinct from the Landau level structure studied for twisted graphene
bilayers in high fields \cite{deGail}. Low angle faults offset the Dirac nodes at the zone corners $K$ in neighboring layers
introducing an energy scale $\Delta E = \hbar v_F K \sin(\theta/2)$ at which layer-decoupled Dirac cones intersect. Interlayer
mixing of these states creates a saddle point singularity where the topology of the bands changes; in a magnetic field this is
identified by an onset of enhanced coherence-splittings of nearly layer-degenerate LL states.   For a rotation angle $\sim
1^{\circ}$, $\Delta E \sim 150 \, {\rm meV}$ so for the $n$-th Landau level this occurs for $nB \sim 30 \, {\rm T}$. Likewise
the physics of the Dirac comb is distinct from the Hofstadter physics that arises when the magnetic length becomes commensurate
with a superlattice translation, requiring $B \sim 4 \, {\rm T}$ for $\theta \sim 1^{\rm \circ}$ \cite{butterfly}.

In summary we have studied the Dirac comb in the weak field Landau quantized spectra for twisted graphene bilayers: an
interference phenomenon yielding an amplitude modulation of the LL spectra. This is important for low energy magnetotransport
and can be used to identify weak interlayer coherence and its low energy mass structure.

This work was supported by the Department of Energy, Office of Basic Energy Sciences under contract DE-FG02-ER45118 (EJM) and
by the NSF DMR-0820382 (MK).

\end{document}